\documentclass[twocolumn,prl,preprintnumbers,amsmath,amssymb,nofootinbib,superscriptaddress]{revtex4}
\usepackage{slashed}
\usepackage{mathtools}
\usepackage{amsfonts}
\usepackage{amssymb}
\usepackage{epsfig}
\usepackage{rotating}
\usepackage{url}
\usepackage{times}
\usepackage{color}
\usepackage{bm}
\usepackage{xcolor,colortbl}
\usepackage{hyperref}
\usepackage[alsoload=hep]{siunitx}
\usepackage{enumitem}
\usepackage{fancyhdr}
\usepackage{anyfontsize}
\usepackage{empheq}

\newcommand{\grad}{\ensuremath{\vec{\nabla}}}

\newcommand{\phib}{{\bar{\phi}}}

\newcommand{\nn}{\nonumber}

\newcommand{\PsiE}{\hat{\Psi}}
\newcommand{\PhiE}{\hat{\Phi}}
\newcommand{\alphat}{\alpha}

\newcommand{\metE}{\ensuremath{\hat{g}}}

\newcommand{\metM}{\ensuremath{g}}

\newcommand{\RiemE}{\ensuremath{\hat{R}}}

\DeclareMathSizes{10}{10}{7}{6}

\definecolor{orange}{rgb}{1,0.5,0}
\definecolor{darkorange}{rgb}{0.69,0.33,0.13}
\definecolor{fidcol}{rgb}{0.7,0,0}

\begin{document}
	
\title{A general class of gravitational theories as alternatives to dark matter where the speed of gravity always equals the speed of light.}

\author{Constantinos Skordis}
\email{skordis@fzu.cz}
\author{Tom~Z\l o\'{s}nik}
\email{zlosnik@fzu.cz}

\affiliation{
		 CEICO, Institute of Physics of the Czech Academy of Sciences, Na Slovance 1999/2, 182 21, Prague 
}

%
%

\begin{abstract}
A number of theories of gravity have been proposed as proxies for dark matter in the regime of galaxies and cosmology.
The recent observations of gravitational waves (GW170817) from the merger of two neutron stars, followed by an electromagnetic counterpart (GW170817a)
have placed stringent constraints on the difference of the speed of gravity to the speed of light,
severely restricting the phenomenological viability of such theories. We revisit the
impact of these observations on the  Tensor-Vector-Scalar (TeVeS) paradigm of relativistic Modified Newtonian Dynamics (MOND) 
and demonstrate the existence of a previously unknown class of this paradigm where the speed of gravity always equals the speed of light.
We show that this holds without altering the usual (bimetric) MOND phenomenology in galaxies. 
\end{abstract}

\maketitle

\paragraph{Introduction}
In the absence of direct detection of a particle with the right properties to account for the entirety of dark matter, it remains a possibility that the effects attributed to dark matter represent a shortcoming in our understanding of the nature of gravity, that is, General Relativity (GR) may not describe gravity correctly in all curvature regimes.
Attempts account for this~\cite{Sanders1997,Bekenstein2004,Skordis2008,Milgrom2009,BabichevDeffayetEsposito-Farese2011} introduce additional fields  into the gravitational sector
whose influence on  the visible matter produces dark matter like effects. In all metric theories of gravity including the ones with additional fields,
 spacetime is a dynamical entity leading to the generation and propagation of gravitational waves. Any additional fields
coupled non-trivially to the spacetime curvature, however, generally lead to gravitational wave speed different than GR.


Gravitational waves (GW) from the merger of a binary neutron star system
have been observed by the advanced Laser Interferometer Gravitational Observatory (aLIGO)
 and the VIRGO interferometer \cite{TheLIGOScientific:2017qsa}. 
Within seconds of this event (GW170817) being detected, a gamma ray burst was independently observed from the same location \cite{Goldstein:2017mmi,Savchenko:2017ffs}. Given the high likelihood that these represent signals from the same event, the specific small time difference - given the large distance between the location of emission (the galaxy NGC 4993) - implies that (in units where the speed of light is unity), the speed of propagation of GW $c_{T}$ obeys
\begin{align}
|c_{T}^{2}-1| \lesssim 10^{-15}.
\label{ct_constraint}
\end{align}
This is a remarkably stringent constraint and has led to many modified theories of gravity proposed in order to explain the phenomenon of dark energy,
 being excluded \cite{BakerEtAl2017, EzquiagaZuma2017,CreminelliVernizzi2017, SaksteinJain2017,CopelandEtAl2018} for such a purpose.
Equally important is the impact of these observations on gravitational theories functioning as effective dark matter proxies.
This stringent constraint has also been used in \cite{OostEtAl2018} to place constraints on the Einstein-Aether theory~\cite{Jacobson2000}.

Early evidence for dark matter came in the form of observations of the motion of stars within galaxies \cite{RubinEtAl1980}, where it was found that stars towards the outer regions of galaxies had orbital velocity significantly higher than expected due to the Newtonian gravitational field produced by visible matter. In 1983, Milgrom showed \cite{Milgrom1983} that this motion of stars could instead result from a modification to the inertia/dynamics of stars at low Newtonian accelerations. 
Shortly afterwards it was found that these same effects could alternatively result from a non-linear modification to the Poisson
 equation of Newtonian gravity \cite{BekensteinMilgrom1984}. These models are referred to as MOdified Newtonian Dynamics (MOND). 

A great deal of work has gone into deducing the astrophysical consequences of MOND, and whether MOND is consistent with data \cite{SandersMcGaugh2002,BinneyFamaey2005,BekensteinMagueijo2006,GentileEtAl2007,MagueijoMozaffari2011,Heesetal2015,MargalitShaviv2015,McGaughEtAl2016,HodsonZhao2017,FrandsenPetersen2018,LelliEtAl2019}. The inherently non-relativistic nature of this modification renders it difficult to test as its realm of validity is unclear. As a result, there have been a number of proposals for relativistic theories that yield MONDian behavior on galactic 
scales \cite{NavarroVanAcoleyen2005,ZlosnikFerreiraStarkman2006,Sanders2007,Milgrom2009,BabichevDeffayetEsposito-Farese2011,Woodard2014,Khoury2014,Berezhiani:2015pia,Berezhiani:2015bqa}. 

Perhaps the most widely-known relativistic theory leading to MOND-like behavior
 is the Bekenstein-Sanders Tensor-Vector-Scalar (TeVeS) theory \cite{Sanders1997,Bekenstein2004}
which depends on a metric $\metE_{\mu\nu}$, a unit-timelike vector field $A_\mu$ and a scalar field $\phi$. 
All types of matter are taken to couple universally to a metric $g_{\mu\nu}$ via
\begin{align}
g_{\mu\nu} &=  e^{-2\phi}\metE_{\mu\nu} - 2\sinh(2\phi)A_{\mu}A_{\nu}
\label{phys_metric}
\end{align}
and as such the Einstein Equivalence Principle is obeyed.
Due to the algebraic relation between the two metrics, there is only one tensor mode propagating gravitational wave perturbation (2 polarizations) in this theory 
just as in GR. The cosmology of TeVeS theory has been extensively investigated in~\cite{SkordisEtAl2005,BourliotEtAl2006,XuEtAl2014,ZlosnikSkordis2017}.

The speed of the tensor mode GW in TeVeS theory is in general different than the speed of light, and
it is then natural to ask what is the status of the TeVeS paradigm after GW170817. Using a variety of methods, 
a number of articles~\cite{BoranEtAl2017,GongEtAl2018,HouGong2018} have tackled this question.
The authors of~\cite{BoranEtAl2017} compared the Shapiro time delay of gravitational versus electromagnetic waves,
 as they pass through the potential wells of galaxies, proposed earlier as a generic test of TeVeS theory~\cite{KahyaWoodard2007}. 
Such a test is superior to testing the propagation speed on a Friedman-Robertson-Walker (FRW) background considered 
in ~\cite{BakerEtAl2017, EzquiagaZuma2017,CreminelliVernizzi2017, SaksteinJain2017,OostEtAl2018} in the case of other theories. 
The delay was calculated there by comparing the geodesics of $\metE_{\mu\nu}$ to the geodesics of $g_{\mu\nu}$, however,
as the metric is not an observable the generality of their result is unclear.
For instance, \cite{ZlosnikFerreiraStarkman2006b} reformulated TeVeS theory using a single metric ($g_{\mu\nu}$) so that
no geodesic comparisons are possible in that formulation.~\footnote{
Other  single-metric theories such as the Horndeski theory studied in~\cite{BakerEtAl2017, EzquiagaZuma2017,CreminelliVernizzi2017, SaksteinJain2017,CopelandEtAl2018}
and Einstein-Aether theory studied in \cite{OostEtAl2018} may also yield Shapiro time delay different than the one in GR.} 
A different method is necessary.

 In~\cite{GongEtAl2018,HouGong2018} the speed of all six types of GW present in TeVeS theory~\cite{Sagi2010} 
has been considered on a Minkowski background and after imposing \eqref{ct_constraint}, analysis of the remaining parameter space led to the conclusion that TeVeS 
theory is ruled out.

In this article we investigate the propagation of GW on perturbed Friedman-Robertson-Walker (FRW) spacetimes, which includes
the Shapiro time delay effect. We find that the original TeVeS theory~\cite{Bekenstein2004} and its generalization~\cite{Skordis2008,Skordis2009} are ruled out by
the GW170817/GW170817a events, in agreement with previous studies~\cite{BoranEtAl2017,GongEtAl2018,HouGong2018}.
We present, however, the existence of a previously unknown class of relativistic MOND theories also based on the Tensor-(timelike)Vector-Scalar paradigm,
where the speed of gravity always equals the speed of light while retaining the effective bi-metric description leading to the usual MOND phenomenology in galaxies. 

\paragraph{Rudiments of TeVeS theory}
A slight generalization of TeVeS is given by the following action~\cite{Skordis2008,Skordis2009}
 which depends on the three above fields and the two auxiliary fields $\lambda_A$ and $\mu$:
\begin{align}
\hat{S} &= \frac{1}{16\pi G} \int d^{4}x \sqrt{-\hat{g}}\bigg[\hat{R}- \hat{K} +\lambda_A \left(A^{\rho}A_{\rho}+1\right)
\nn
\\
&
-\mu \hat{g}^{\alpha\beta}\hat{\nabla}_{\alpha}\phi\hat{\nabla}_{\beta}\phi-\hat{V}(\mu)
\bigg] + S_{{\rm M}}[g]
 \label{teves_action}
\end{align}
Here $G$ is the bare gravitational constant, $\hat{g}$ and  $\RiemE$ are the determinant and scalar curvature of $\hat{g}_{\mu\nu}$ respectively,
$\hat{V}$ is a free function of $\mu$, $S_M[g]$ is the action for all matter fields 
and $\hat{K}= \hat{K}^{\mu\nu\alpha\beta}\hat{\nabla}_{\mu}A_{\nu}\hat{\nabla}_{\alpha}A_{\beta}$ is obtained using
\begin{equation}
\hat{K}^{\mu\nu\alpha\beta} = 
 c_1\hat{g}^{\mu\alpha}\hat{g}^{\nu\beta}
 + c_2\hat{g}^{\mu\nu}\hat{g}^{\alpha\beta}
+c_3 \hat{g}^{\mu\beta}\hat{g}^{\nu\alpha}
+ c_4\hat{g}^{\nu\beta}A^{\mu}A^{\alpha}
\end{equation}
The indices of $A_\mu$ are always raised using $ \hat{g}^{\mu\nu}$, the inverse metric of $\hat{g}_{\mu\nu}$, i.e. $\hat{g}^{\mu\rho} \hat{g}_{\rho\nu} = \delta^\mu_{\;\;\nu}$). 
We emphasise that in contrast to~\cite{Skordis2008,Skordis2009}  we allow here the $c_{I}$ ($I= 1\ldots4$) to be functions of the scalar field $\phi$
and this comes out to be very important when analyzing the speed of GW.
The original TeVeS theory is obtained 
when $c_I = \{2K_B - \frac{1}{4},  - \frac{1}{2}, -2K_B + \frac{3}{4},  K_B - \frac{1}{4}\}$,
for a constant $K_{B}$~\cite{Skordis2009}. For notational compactness we define $c_{IJ\dots} \equiv c_{I}+c_{J}+\dots$.

The emergence of MOND behavior in the quasistatic weak field limit in the constant $c_I$ case has been analysed extensively in  \cite{Skordis2009}.
We revisit that analysis here in order to show that it remains unchanged even when $c_I $ are functions of $\phi$.
In particular, one expands the scalar field as  $\phi =\phi_{0}+\varphi$ with $\phi_0$ a constant and $\varphi$ time independent.
The quasistatic metric is such that $\hat{g}_{00} = -e^{-2\phi_0}(1-2\PsiE)$ and $\hat{g}_{ij} = e^{2\phi_0}(1-2\PhiE)\gamma_{ij}$. 
In this  coordinate system the vector field has components $A_0 = -e^{-\phi_0}(1+\PsiE)$ and $A_i =0$. 
Using the metric transformation  \eqref{phys_metric} we find the components of the
metric $g_{\mu\nu}$  so that
\begin{equation}
ds^{2} =  -(1+2\Psi)dt^{2} + (1-2\Phi)\gamma_{ij}dx^{i}dx^{j}
\label{quasistatic_metric}
\end{equation}
 where
\begin{equation}
\PsiE = \Psi - \varphi,
\qquad
\PhiE = \Phi - \varphi,
\label{matter_potentials}
\end{equation}
With this ansatz, the vector field equations are identically satisfied while the Einstein and scalar field equations reduce to 
\begin{align}
\vec{\nabla}^{2}\hat{\Psi} &= \frac{8\pi G}{2 -  c_1 + c_4}\rho 
 \label{newmod}
\\
\vec{\nabla}_{i}\big(\mu \vec{\nabla}^{i}\varphi\big) &= 8\pi G\rho 
\\
\hat{\Phi} &= \hat{\Psi}
\label{AQUAL_equations}
\end{align}
where $\rho$ is the matter energy density and where the $c_I$'s are  evaluated at $\phi=\phi_0$ in (\ref{newmod}). 
The non-dynamical field $\mu$ is obtained via a constraint equation found from the action upon variation wrt $\mu$ and this equation depends on the form of $\hat{V}(\mu)$.
Not all functions  $\hat{V}(\mu)$ lead to either Newtonian or MONDian limiting behaviors and the ones that do so must have appropriate properties
 discussed in \cite{Skordis2009}.

\paragraph{Tensor mode propagation on FRW backgrounds}
In order to determine the speed of propagation of GW, we need the tensor mode equation on an FRW background. We assume a metric $g_{\mu\nu}$ such that
\begin{align}
ds^2 =&  - dt^2  +  a^2 \left(\gamma_{ij} + \chi_{ij}\right) dx^i dx^j
\end{align}
where $a$ is the scale factor, $\gamma_{ij}$ is the spatial metric of constant curvature $\kappa$ and $\chi_{ij}$ is the tensor mode GW which is traceless $\gamma^{ij} \chi_{ij}=0$ 
and transverse $\grad_i \chi^i_{\;\;j}=0$, where $\grad_i$ is the spatial covariant derivative compatible with $\gamma_{ij}$. As we are interested only in the tensor mode,
we let the perturbations of $\phi$ and $A_\mu$ to zero so that $\phi = \phib(t)$ and $A_0 = -e^{-\phib}$ with $A_i=0$.
The perturbed Einstein equations for the tensor mode have been obtained for constant $c_I$ in \cite{Skordis2008}. In the case where $c_I = c_I(\phi)$, we find an additional
term present such that
\begin{align}
& 
  e^{2\phib} \left(1 - c_{13} \right) \left[ \ddot{\chi}^i_{\;\;j} + \left(3 H + 4\dot{\phib}\right) \dot{\chi}^i_{\;\;j} \right]
- e^{2\phib} \frac{dc_{13}}{d\phi}  \dot{\phib}  \dot{\chi}^i_{\;\;j} 
\nonumber
\\ 
&
-\frac{1}{a^2} e^{-2\phib} \left(\grad^2 - 2\kappa\right) \chi^i_{\;\;j}
= 
16 \pi G e^{-2\phib} \Sigma^{(g)i}_{\;\;\;\;\;\;j}
\label{GW_FRW}
\end{align}
where $\Sigma^{(g)i}_{\;\;\;\;\;\;j}$ is a traceless source term due to matter.
The only difference from the constant $c_I$ case is the appearance of the $ \frac{dc_{13}}{d\phi}  $ term multiplying $ \dot{\chi}^i_{\;\;j}$.

Now in the original and in the generalized TeVeS theories it is clear that the speed of propagation of the tensor mode is given by
\begin{equation}
c^2_T = \frac{ e^{-4\phib}}{1 - c_{13} }
\end{equation}
Thus, in general $c^2_T$ will differ from unity putting this theory in conflict with the observations that require $c_T^2\approx 1$ unless
 some mechanism sets $\phib$ to be an approximately constant value at very low redshift and equal to $\phib = -\frac{1}{4}\ln(1-c_{13})$.
This is highly unlikely but even if possible, we show below that the Shapiro time delay rules this case out.

If $c_I$ are functions of $\phi$, however, there seems to be enough freedom to change this fact. In particular, the unique choice of 
\begin{align}
c_{13}(\phi) =  1-e^{-4\phi} 
\label{cteq1}
\end{align}
transforms \eqref{GW_FRW} into
\begin{align}
& 
 \ddot{\chi}^i_{\;\;j} + 3 H \dot{\chi}^i_{\;\;j} -\frac{1}{a^2} \left(\grad^2 - 2\kappa\right) \chi^i_{\;\;j}
= 
16 \pi G   \Sigma^{(g)i}_{\;\;\;\;\;\;j}
\label{GW_GR_FRW}
\end{align}
which is identical to the tensor mode equation in GR and thus with this choice $c^2_T = 1$.

\paragraph{Beyond FRW}
We have shown above that the choice \eqref{cteq1} leads to GW tensor mode propagation as in GR while maintaining MONDian behavior.
When gravitational and electromagnetic wave pass through potential wells generated by matter, however, they incur an additional (Shapiro) time delay
 and as \cite{KahyaWoodard2007} proposed, it may be used to put strong constraints on such theories.
We thus examine whether the condition \eqref{cteq1} is sufficient to ensure tensor mode propagation with $c^2_T = 1$ even when including the effect
of inhomogeneities. In these situations the physical metric $g_{\mu\nu}$ takes the form
\begin{align}
ds^{2} &=  -(1+2\Psi)dt^{2} + a^2(1-2\Phi)\left(\gamma_{ij}+ \chi_{ij}\right)dx^{i}dx^{j}
\label{Physical_metric}
\end{align}
where the hierarchy $\chi_{ij} \ll \Phi,\Psi \sim 10^{-5}$ has been assumed. 
Furthermore TeVeS's scalar field $\phi$ takes the form $\phi = \phib +\varphi$ (with $\varphi \ll 1$) while the vector field has components
$A_0 = - e^{-\phib} (1 + \PsiE)$ and $A_i =  - a e^{\phib}  \grad_i \alphat$.  The expressions \eqref{matter_potentials}  relate the potentials 
between the two frames. Given \eqref{Physical_metric} the metric $\hat{g}_{\mu\nu}$ will not be in diagonal form but will contain terms 
coming from the vector perturbation $\alphat$. 
In general the potentials are assumed to be space and time dependent.

Defining ${\cal T}^{ik}_{\phantom{ki}\,j} = \grad^i \chi^k_{\;\;j} + \grad_j \chi^{ki} - \frac{2}{3} \grad^l \chi^k_{\;\;l} \delta^i_{\;\;j}$,
after a lengthy and tedious calculation, the tensor mode equation for $\chi_{ij}$ is found to be 
\begin{widetext}
\begin{align}
& 
  e^{2\phib}  \left[ \left(1  - c_{13} \right)  \left(1 - 2 \PsiE\right)  -\frac{dc_{13}}{d\phi}  \varphi   \right]  \ddot{\chi}^i_{\;\;j} 
+  e^{2\phib} {\cal A} \dot{\chi}^i_{\;\;j} 
-\left\{  \frac{1}{a^2} e^{-2\phib} 
 \left[ (1+2\PhiE) \left(\grad^2 - 2\kappa\right)    + \grad_k(\PsiE - \PhiE ) \grad^k  \right]                                                                                 
-  \frac{1}{a} {\cal B}_k \grad^k\right\} \chi^i_{\;\;j}
\nonumber
\\ 
&
+ \frac{1}{a^2} e^{-2\phib}   (1+2\PhiE) \left(  \grad^i \grad_k \chi^k_{\;\;j} + \grad_j \grad^k\chi^i_{\;\;k} 
- \frac{2}{3}   \grad^l \grad_k \chi^k_{\;\;l} \delta^i_{\;\;j}
\right)
+  \frac{1}{a^2} e^{-2\phib} \bigg\{
  \grad_k(\PsiE - \PhiE ) {\cal T}^{ik}_{\phantom{ki}\,j} 
\nonumber
\\ 
&
- 2  \left[\grad_k\grad_j\left(\PhiE - \PsiE  \right)\chi^{ik} - \frac{1}{3} \grad_k\grad_l\left(\PhiE - \PsiE  \right)\chi^{lk}  \delta^i_{\;\;j} \right]
\bigg\}
+ \frac{1}{a} {\cal C}^i_{\;\;j}                                                                                                                                      
= 
16 \pi G  e^{-2\phib}(1-2\varphi)\Sigma^{(g)i}_{\;\;\;\;\;\;j}
\label{GW_pert}
\end{align}
\end{widetext}
where
the terms ${\cal A}$, ${\cal B}_i$, ${\cal C}^i_{\;\;j}$ are shown in the appendix for clarity of presentation.
Allowing all potentials as well as $\varphi$ and $\alphat$ to vanish reduces \eqref{GW_pert} to \eqref{GW_FRW}.

Consider first the reduction of \eqref{GW_pert} to quasistatic backgrounds, also ignoring the source term,
for the fine-tuned case where $c_{13} = 1 - e^{-4\phi_0}$ (so that $c_T^2=1$ on the background).
We obtain this by setting $a=1$, $\phib=\phi_0$  and $\hat{\Psi}=\hat{\Phi}$ from \eqref{AQUAL_equations}. 
Then ${\cal A}$, ${\cal B}_k$, ${\cal C}^i_{\;\;j}$ all vanish. In addition, considering LIGO wavelengths $\sim1000km$ which are far smaller
than the scale of the potential wells, we may drop the terms containing derivatives on $\Psi$ and $\varphi$, i.e $\partial\Phi \ll \partial \chi$,
even with $\chi\ll \Phi$ \cite{CopelandEtAl2018}.
Finally, imposing further the gauge condition $\grad_i\chi^i_{\;\;j}=0$, \eqref{GW_pert} leads to
\begin{align}
& 
 \left(1 - 2 \PhiE\right)  \ddot{\chi}^i_{\;\;j} -  \left(1+2\PhiE\right) \grad^2    \chi^i_{\;\;j} = 0 
\end{align}
Thus in this case we expect a Shapiro time delay dictated by $\PhiE$, the potential formed by baryons alone. This is not the same as $\Phi$ which is the potential seen by photons, hence, this fine-tuned case is ruled out by the analysis of~\cite{BoranEtAl2017}.

Let us turn now to the case where \eqref{cteq1} holds so that $c_T^2=1$ on FRW.
Imposing \eqref{cteq1} we find ${\cal B}_k = 0$ and ${\cal C}^i_{\;\;j} =0$.
Further using  \eqref{matter_potentials} we find $e^{2\phib}{\cal A} =  e^{-2\phib}\left[ 3 H (1 - 2 \Psi - 2 \varphi) - \dot{\Psi} - 3 \dot{\Phi}\right]$
 and after choosing the gauge condition  $\grad_i\chi^i_{\;\;j}= \grad_i (\Phi -\Psi)\chi^i_{\;\;j}$, \eqref{GW_pert}  turns into
\begin{align}
& 
 \left(1 - 2 \Psi  \right)  \left[ \ddot{\chi}^i_{\;\;j} +  \left( 3 H  - \dot{\Psi} - 3 \dot{\Phi} \right)\dot{\chi}^i_{\;\;j}  \right]
-  \frac{1}{a^2} \left(1+2\Phi\right) \bigg[
\nonumber
\\
&
  \left(\grad^2 - 2\kappa\right) \chi^i_{\;\;j} 
 +\grad_j   \grad^k (\Phi -\Psi) \chi^i_{\;\;k} 
 - \grad^i   \grad_k (\Phi -\Psi) \chi^k_{\;\;j}
\nonumber
\\
&
 - \grad_k(\Phi - \Psi )  \grad^k \chi^i_{\;\;j} 
\bigg]
= 
16 \pi G  \Sigma^{(g)i}_{\;\;\;\;\;\;j}
\label{GW_GR_pert}
	\end{align}
which is the same equation as in GR.  Thus, with the choice \eqref{cteq1}  the tensor mode propagates at the speed of light even when including inhomogeneities
and gives the same Shapiro time delay as for photons.

\paragraph{Tensor mode propagation on general backgrounds}
Our results of the previous section are not an accident. To gain further insight as to why this behavior emerges we consider 
the single-metric (physically equivalent) formulation of TeVeS. As shown in~\cite{ZlosnikFerreiraStarkman2006b}, 
one introduces a new field $B_\mu = A_\mu$ leading to $B^\mu = e^{-2\phi} A^\mu$
and writes the Lagrange constraint  as $g^{\mu\nu}B_\mu B_\nu \equiv B^2 =  -e^{-2\phi}$.
This enables us to solve for $\phi$ and thus remove both $\phi$ and $\hat{g}_{\mu\nu}$ from the action.  The dof remain unchanged as now 
$B_\mu$ contains 4 dof rather than the 3 of $A_\mu$. The action $S[g,B,\mu]$ of this physically equivalent Vector-Tensor formulation is
\begin{align}
S =  \frac{1}{16\pi G} \int d^4x \sqrt{-\metM} \left[ R  - K - U\right] + S_m[g]
 \label{single_frame_teves_action}
\end{align}
where $U = \hat{V}(\mu)/B^2$ and $K$ is given by
\begin{align}
K &=   (d_1-d_3) F^{\mu\nu}F_{\mu\nu} + d_{13} M^{\mu\nu}M_{\mu\nu} 
+ d_2 J^2  + d_4 J^\nu J_\nu 
\nn
\\
&
+ \frac{1}{2} d_5  J^\mu\nabla_\mu B^2 + \frac{d_6}{4} (\nabla B^2)^2
+ \frac{d_7}{2} Q J + \frac{d_8}{4}Q^2
 \label{K_tensor}
\end{align}
and we have also defined $F_{\mu\nu} =2\nabla_{[\mu} B_{\nu]}$, $M_{\mu\nu} =2\nabla_{(\mu} B_{\nu)}$,
$J = \nabla_\mu B^\mu$, $J_\mu = B^\alpha \nabla_\alpha B_\mu$ and $ Q =  B^\alpha  \nabla_\alpha B^2$.
The functions for the generalised TeVeS theory $d_I$ ($I=1\ldots 8$)  may be found in the appendix of \cite{Skordis2009}.

 Some of the $d_I$ coefficients depend on $\mu$
so that MOND behavior may emerge upon choosing appropriate $\hat{V}$. Allowing for a general dependence $d_{I}(B^{2},\mu)$ in (\ref{single_frame_teves_action})
 represents a slight generalization of (\ref{teves_action}).
 Interestingly, the dynamical tendency towards $B_\mu$ having a non-vanishing norm in this picture arises from 
the presence of inverse powers of the norm $B^{2}$ in the Lagrangian, rather than via a Lagrangian constraint as in (\ref{teves_action}).
 In this formulation, the modification to the speed of propagation of GW is due entirely to the the coupling of gravity to the field $B_\mu$
 through that field's kinetic term. A straightforward way to see this is by considering the case where $B^\mu$ is hypersurface orthogonal;
 in which case we can decompose the metric $g_{\mu\nu}$ as $g_{\mu\nu} = h_{\mu\nu} - n_{\mu}n_{\nu}$
where $n_\mu \equiv B_\mu/\sqrt{-B^2} = N\nabla_\mu t$ for some global time function $t$ and $h_{\mu\nu}$ ($h_{\mu\nu}n^{\nu}=0$) is the spatial metric on surfaces of constant time. Then
\begin{align}
K &=  -d_{13}B^{2}{\cal K}^{\mu\nu}{\cal K}_{\mu\nu}-d_{2}B^{2}{\cal K}^{2} + \dots \label{kineticus}
\end{align}
where we have defined the extrinsic curvature tensor ${\cal K}_{\mu\nu} \equiv \frac{1}{2}{\cal L}_{n} h_{\mu\nu}$, 
and $\dots$ denote terms of linear order or lower in ${\cal K}_{\mu\nu}$. 
As gravitational wave perturbations reside in `trace-free' small perturbations to $h_{\mu\nu}$, only the first term in (\ref{kineticus}), schematically of the form 
$\sim d_{13} B^{2}\dot{h}^{\mu\nu}\dot{h}_{\mu\nu}$ will affect the speed of gravity. There will be no deviation from General Relativity if 
\begin{equation}
d_{13} = 0 \qquad \Rightarrow \quad d_1 = -d_{3}. 
\end{equation}
The transformation of \eqref{teves_action} 
into \eqref{single_frame_teves_action} gives $d_{13} = \frac{1 - c_{13}  }{B^6} - \frac{1}{B^2}$ so that $d_{13}=0$ iff $c_{13} = 1-B^{4} =1-e^{-4\phi}$,
which is condition \eqref{cteq1}.

\paragraph{Discussion and conclusions}
We have demonstrated the existence of a  generic class of relativistic theories of MOND based on the Tensor-(timelike)Vector-Scalar paradigm 
which retain the property that GW in this class propagate as in General Relativity.
The original TeVeS theory is not part of this class and therefore not consistent with gravitational wave constraints.
Hence, actions of the form (\ref{single_frame_teves_action}) are sufficiently general that they encompass both phenomenologically viable and non-viable models. 
Viable models are those for which $d_3 = - d_1$ so that the $M_{\mu\nu}$ term is absent while all remaining $d_I$'s can in general be functions of both $B^2$ and $\mu$.
However, not all such viable actions lead to MOND behavior but specific functional forms of $d_I$ do so. Indeed, it is possible 
to sufficiently simplify the viable subset of (\ref{single_frame_teves_action}) while retaining a MOND limit and at the same time giving  a realistic cosmology. 
This is beyond the scope of this article and will be investigated elsewhere.

\begin{acknowledgments}
We thank M. K. Kopp and R. P. Woodard for discussions. 
We further thank M. K. Kopp for teaching us how to use the xPand~\cite{PitrouRoyUmeh2013} Mathematica package with which our analytical calculations were cross-checked.
 The research leading to these results has received funding from the European Research Council under the European Union's 
Seventh Framework Programme (FP7/2007-2013) / ERC Grant Agreement n. 617656 ``Theories
 and Models of the Dark Sector: Dark Matter, Dark Energy and Gravity'' and from the European Structural and Investment Funds
and the Czech Ministry of Education, Youth and Sports (MSMT) (Project CoGraDS - CZ.02.1.01/0.0/0.0/15003/0000437).   
\end{acknowledgments}

\appendix

\begin{widetext}
\section{Coefficients of the gravitational wave equation}
\label{Various_GW_tensors}

Defining
${\cal E}^i_{\;\;j} = \grad^i  \alphat   \grad_k  \dot{\chi}^k_{\;\;j} -  \frac{1}{3}  \grad^l  \alphat   \grad_k  \dot{\chi}^k_{\;\;l} \delta^i_{\;\;j}$
the tensors ${\cal A}$, ${\cal B}_i$ and ${\cal C}^i_{\;j} $ are given by
\begin{align}
 {\cal A} &= \left(1 - c_{13} \right) \left[ (3 H + 4\dot{\phib})  \left(1 - 2 \PsiE\right)  - \dot{\PsiE} - 3 \dot{\PhiE} \right]
-  \frac{dc_{13}}{d\phi} \left[ (3 H +4 \dot{\phib})   \varphi    + \dot{\phib}   (1 -  2\PsiE) +  \dot{\varphi } \right]
 - \frac{d^2c_{13}}{d\phi^2} \dot{\phib}  \varphi  
\nonumber
\\
&
+  \frac{1}{a}   e^{-2\phib}\left( 1- e^{4\phib} + e^{4\phib}  c_{13}  \right)  \grad^2  \alphat 
-2  \left(e^{4\phib}  - 1 -  e^{4\phib} c_{13} \right) \grad_k  \alphat  \grad^k 
\\
{\cal B}_i &=  -  \left(e^{4\phib}  - 1 -  e^{4\phib} c_{13} \right)  \grad_i  \dot{\alphat} 
- 2  \bigg\{ \left[e^{4\phib}\left(1 - c_{13} \right)  - 1  \right]  H 
+ \left[3e^{4\phib}\left(1 - c_{13} \right)  - 1  - \frac{1}{2} e^{4\phib} \frac{dc_{13}}{d\phi} \right] \dot{\phib} 
   \bigg\} \grad_i \alphat  
\\
{\cal C}^i_{\;j} 
&= 
  \left(e^{4\phib}  - 1 -  e^{4\phib} c_{13} \right) \bigg\{
   \grad_k \grad_j \alphat \dot{\chi}^{ik} 
 + \grad_k \grad^i \alphat \dot{\chi}^k_{\;\;j}
 - \frac{2}{3}  \grad_k \grad_l \alphat \dot{\chi}^{lk}  \delta^i_{\;\;j} 
+ 2\left( \grad_k\grad_j \dot{\alphat} \chi^{ik} - \frac{1}{3}  \grad_k\grad_l \dot{\alphat} \chi^{lk}   \delta^i_{\;\;j} \right)
\nonumber
\\
&
 + \grad_k  \dot{\alphat}  {\cal T}^{ik}_{\phantom{ki}\,j}
+ \grad_k  \alphat   \dot{{\cal T}}^{ik}_{\phantom{ki}\,j} 
\bigg\}
 +   \left(e^{4\phib}  - 1 - \frac{1}{2} e^{4\phib} c_{13} \right) {\cal E}^i_{\;\;j} - \frac{1}{2} c_{13} {\cal E}_j^{\;\;i} 
+ 2   \bigg\{ \left[e^{4\phib}\left(1 - c_{13} \right)  - 1  \right]  H 
 \nonumber \\ 
& 
+ \left[3e^{4\phib}\left(1 - c_{13} \right)  - 1  - \frac{1}{2} e^{4\phib} \frac{dc_{13}}{d\phi} \right] \dot{\phib} 
   \bigg\} \left[ 
\grad_k \alphat {\cal T}^{ik}_{\phantom{ki}\,j} 
+  2 \left(
 \grad_k\grad_j \alphat \chi^{ik} -   \frac{1}{3} \grad_k\grad_l \alphat \chi^{lk} \delta^i_{\;\;j}\right) \right]
\end{align}
\end{widetext}


%
%
%
%

\bibliographystyle{unsrtnat}
\bibliography{tzreferences}

\end{document}